\documentstyle[preprint,aps]{revtex}
\tightenlines

\begin{document}

\draft
\title
{Topological Invariants and Anyonic Propagators}

\author
{Wellington da Cruz\footnote{E-mail: wdacruz@fisica.uel.br}}

\address
{Departamento de F\'{\i}sica,\\
 Universidade Estadual de Londrina, Caixa Postal 6001,\\
Cep 86051-970 Londrina, PR, Brazil\\}
 
\date{\today}

\maketitle

\begin{abstract}
We  obtain the Hausdorff dimension, $h=2-2s$, for particles
with fractional
spins in the interval, $0\leq s \leq 0.5$, such that 
the manifold
is characterized by a topological invariant given by,
${\cal W}=h+2s-2p$.
This object is related to fractal properties of the
path swept out by fractional spin 
particles, the spin of these particles, and the
genus ( number of anyons ) of 
the manifold. We prove that the anyonic propagator
can be put into a path 
integral representation which gives us a continuous
family of Lagrangians 
in a convenient gauge. The formulas for, $h$ and ${\cal W}$,
were obtained 
taking into account the anyon model as a particle-flux
system and by 
a qualitative inference of the topology.  
\end{abstract}

%\narrowtext

%\

In this letter, we obtain from anyonic propagators
 the Hausdorff
dimension for fractional spin particles with spin
 in the interval,
$0\leq s \leq 0.5$, that is, between scalar and
 spinning particles.
 These 
fractal properties of the path swept out by such particles 
is given by,
 $h=2-2s$. 
On the other hand, we have that the Euler number is a
 topological 
invariant which characterize closed surfaces with genus $p$,
 that is, ${\cal E}=2-2p$. In our context we consider
  $p$ the number
 of particles 
with fractional spins on the manifold, such
 that the anyon
model as a particle-flux system implies 
multiply connected spaces. In this way,
 we propose a 
topological invariant given by, ${\cal W}=h+2s-2p$,
 where $h$ is the
Hausdorff dimension of the trajectory, $s$ is the spin
 of the particle and 
$p$ is the number of fractional spin particles ( holes ) of
 the manifold.
Formulas for $h$ and ${\cal W}$, therefore, characterize the
 particles and the
space where they exist. We have that, for $p=0$,$\;
$${\cal W}=2$, the 
scalar particle, $s=0$, gives $h=2$ and the spinning particle,
 $s=0.5$, gives 
$h=1$. And for $s$, interpolating between these two
 extremes, the number 
$p$ is any, $p=1,2,3,...,N$. In this way, we have
 in fact particles 
with fractional spins. Now, we prove that the propagator 
given in\cite{R1}  

\begin{equation}
\label{f1}
{\tilde{\cal F}}(p)=\frac{1}{(p_{\mu}-mS_{\mu})^2-m^2},
\end{equation}
 
\noindent  where $ m $ and $\alpha$ are 
respectively the mass and the helicity of the anyon, 
and $ s$ is an arbitrary constant deduced from the 
Casimir of the spin algebra, $ S^2=-s^2$ and 
$ S_{\mu}\;p^{\mu}+\alpha m=0$; 
has a path integral representation, 
such that we extract a continuous family of Lagrangians

\begin{equation}
\label{b1}
L_{s}=\frac{{\dot x}^2}{2e}+\frac{s^2}{2e}
+\frac{e}{2}(m^2-2m^2\alpha).
\end{equation}

We consider ${\cal F}^c(x,y)$ as a causal Green function
 for the equation

\begin{equation}
\label{a1}
(P^2-\zeta^2+i\epsilon){\cal F}^c(x,y)=-\delta^3(x-y),
\end{equation}

\noindent where $ P_{\mu}=i\partial_{\mu}-gA_{\mu}(x)$,$\;$ 
$\zeta^2=m^2s^2+(m^2-2m^2\alpha)$ with $ m $ and $\alpha$,
respectively, the mass and
the helicity of the fractional spin particle, $g$ is the
 electric charge and $s$ is an arbitrary constant 
 extracted from the Casimir algebra, $S^2=-s^2$. 
Following\cite{R2} the function ${\cal F}^c(x,y)$ is 
considered as a
matrix element of an operator ${\cal F}^c$

\begin{equation}
{\cal F}^c(x,y)=\langle x|{\cal F}^c|y \rangle,
\end{equation}

\noindent where $ |x\rangle $ are eigenvectors for
 the operators
 $X^{\mu}$ and $P_{\mu}$ is the corresponding canonical
  operator 
 for the momenta $ p_{\mu}$. At this stage, we assume
  some usual
relations.
 
 Eq.(\ref{a1}) implies the equation for the operator
 
\begin{equation}
{\cal F}^c =\left(\zeta^2-\Pi^2-\imath\epsilon\right)^{-1},
\end{equation}

\noindent where $\Pi_{\mu}=-P_{\mu}-g A_{\mu}(x)$ 
and the inverse of the anyonic operator
 is put in the form

\begin{equation}
{\cal F}^c=\imath\int_{0}^{\infty}\exp
\left\{-\imath{\cal H}(\lambda,X,P)\right\}\;
d\lambda,
\end{equation}

\noindent with ${\cal H}(\lambda,X,P)=
\lambda\left(\zeta^2-\Pi^2-\imath\epsilon\right)$.

In this way, we can write

\begin{equation}
\label{a2}
{\cal F}^c(x_{out},x_{in})=\imath\int_{0}^{\infty}
\langle x_{out}
 |e^{-\imath
{\cal H}(\lambda,X,P)} |x_{in}\rangle \;d\lambda.
\end{equation}

The matrix element Eq.(\ref{a2}), taking 
into account the
 Weyl ordering procedure for operators,
 appears now as a path
 integral in
 the phase space:
  
\begin{eqnarray}
&&{\cal F}^{c}(x_{out},x_{in})=\imath\int_{0}^{\infty}
\;d\lambda_{0}
\lim_{N\rightarrow\infty}\int_{-\infty}^{\infty}
\;dx_{1}\cdots dx_{N-1}
\;d\lambda_{1}\cdots d\lambda_{N}\\
&&\times\langle x_{out}|e^{-\imath{\cal H}
(\lambda_{N},X,P)/N}|x_{N-1}\rangle\delta
(\lambda_{N}-\lambda_{N-1})
\cdots\nonumber\\
&&\times\langle x_{i} |e^{-{\cal H}
(\lambda_{i},X,P)/N} |x_{i-1}\rangle
\delta(\lambda_{i}-\lambda_{i-1})\cdots\nonumber\\
&&\times\langle x_{1} |e^{-\imath{\cal H}
(\lambda_{1},X,P)/N} |x_{in}
\rangle\delta(\lambda_{1}-\lambda_{0})
\nonumber\\
&&=\imath\int_{0}^{\infty}\;d\lambda_{0}
\lim_{N\rightarrow\infty}
\int_{-\infty}^{\infty}
dx_{1}\cdots dx_{N-1}\frac{dp_{1}}{(2\pi)^3}
\cdots\;
\frac{dp_{N}}
{(2\pi)^3}\;d\lambda_{1}\cdots
\;d\lambda_{N}\\
&&\times\frac{d\pi_{1}}{2\pi}\cdots
\frac{d\pi_{N}}{2\pi}
\exp\left\{\imath\sum_{k=1}^{N}\left[p_{k}
\frac{\Delta {x_{k}}}{\Delta\tau}-
{\cal H}\left(\lambda_{k},{\bar x}_{k},p_{k}
\right)+\pi_{k}
\frac{\Delta\lambda_{k}}
{\Delta\tau}\right]{\Delta\tau}\right\},\nonumber
\end{eqnarray}

\noindent where ${\Delta{\tau}}=\frac{1}{N}$,
$\;$${\bar{x}_{k}}
=\frac{x_{k}+x_{k-1}}{2}$,$\;$
${\Delta {x_{k}}}=x_{k}-x_{k-1},$$\;$$
{\Delta {\lambda_{k}}}=\lambda_{k}-\lambda_{k-1}$.

 Thus, with $e=2\lambda$ and the 
 shift $-p_{\mu}\rightarrow
 \frac{{\dot x}}{2\lambda}+g\;
 {\dot x}\;.\;A(x)$, we have
 
\begin{eqnarray}
\label{p1}
&&{\cal F}^c(x_{out},x_{in})=\frac{\imath}{2}
\int_{0}^{\infty} 
de_{0}\int_{x_{in}}^{x_{out}}
{\cal D}x\;{\cal D}e\;{\cal D}\pi\;\;M(e)\\
&&\times\exp\left\{-\imath\int_{0}^{1}
\left[\frac{{\dot x}^2}{2e}+\frac
{e\zeta^2}{2}+g\;{\dot x}\;.\;A(x)+\pi{\dot e}\right]\;
d\tau\right\}\nonumber\\
&&=\frac{\imath}{2}\int_{0}^{\infty}\;de_{0}
\;\int{\cal D}x
\;{\cal D}e\;{\cal D}\pi\;
M(e)\\
&&\times\exp\left\{-\imath\int_{0}^{\infty}
\left[\frac{{\dot x}^2}{2e}+\frac{em^2s^2}{2}
+\frac{e{\cal M}^2}{2}
+g\;{\dot x}\;.\;A(x)+\pi{\dot e}\right]
d\tau\right\},\nonumber
\end{eqnarray}

\noindent where we define the measure

\begin{eqnarray}
M(e)=\int {\cal D}p\;\exp\left\{\frac{\imath}{2}
\int_{0}^{1}e\;p^2\; 
d\tau\right\},
\end{eqnarray}

\noindent and ${\cal M}^2=m^2-2m^2\alpha$.

In the gauge ${\dot e}=0$, we obtain

\begin{eqnarray}
&&{\cal F}^c(x_{out},x_{in})=\frac{\imath}{2}
\int_{0}^{\infty} 
de_{0}\int_{x_{in}}^{x_{out}}
{\cal D}x\;{\cal D}e\;{\cal D}\pi\; M(e)\;\delta({\dot e})\\
&&\times\exp\left\{-\imath\int_{0}^{1}
\left[\frac{{\dot x}^2}{2e}+
\frac{s^2}{2e}+\frac{e{\cal M}^2}{2}+g\;{\dot x}\;.\;A(x)
+\pi{\dot e}\right]
d\tau\right\},\nonumber
\end{eqnarray}

\noindent where we recognize the continuous
 family of Lagrangians Eq.(\ref{b1})
 and a gauge-fixing term.
 
 For a free anyon, we have 

\begin{eqnarray}
\label{p1}
{\cal F}^c(x_{out},x_{in})=&&\frac{\imath}{2}
\int_{0}^{\infty} 
de_{0}\int_{x_{in}}^{x_{out}}
{\cal D}x\;{\cal D}e\; M(e)\;\delta({\dot e})\\
&&\times\exp\left\{-\frac{i}{2}\int_{0}^{1}
\left[\frac{{\dot x}^2}{e}+
e{\cal M}^2+\frac{s^2}{e}\right]
d\tau\right\},\nonumber
\end{eqnarray}
 
\noindent such that, making the substitution

\[
{\sqrt {e}}p\rightarrow p, \; 
\frac{x-x_{in}-\tau{\Delta x}}{\sqrt e} \rightarrow x,\;
{\Delta x}=x_{out}-x_{in},\nonumber
\]
\noindent we obtain the new boundary conditions
 $x(0)=0=x(1)$. Thus, we arrive at the path
  integral representation for the propagator
\begin{eqnarray}
\label{k1}
&& {\cal F}^c(x_{out},x_{in})=\sqrt{\frac{\imath}
{4(2\pi)^3}}
\int_{0}^{\infty}
\frac{de_{0}}{\sqrt{e_{0}^3}}\exp\left\{-\frac{\imath}{2}
\left[e_{0}
{\cal M}^2+\frac{{\Delta x}^2}{e_{0}}-\frac{s^2}
{e_{0}}\right]
\right\}
\end{eqnarray}

\noindent and $s^2>0$ for the causal propagator
\cite{R1}. On the other hand, in momentum space,
 the anyonic propagator is given byEq.(\ref{f1}) 
 and for an anyon interacting with an external 
 field, we have a candidate for an equation of motion:
 
 \begin{equation}
 \left(P^2-\zeta^2\right){\cal Y}=0,
 \end{equation}
 
 \noindent where ${\cal Y}$ represents an anyon.
 
 Now, we return to analysis of the Hausdorff dimension
 of the anyonic system. Our inference of the form
  of $h$ takes 
 into account that, for $L$, the length of closed
  trajectory
 with size $R$, we have that its fractal properties
  is given by 
 $L\sim R^h$\cite{R3}, thus for scalar particle 
 $L\sim \frac{1}{p^2}$, $R^2\sim L$ and $h=2$ ; for  
 spinning particle, $
 L\sim \frac{1}{p}$, $R^1\sim L$ and $h=1$. For 
 fractional spin 
 particles, the propagator as given in Eq.(\ref{f1})
 has an explicit 
 dependence on $s$, so we arrive at the formula,
  $h=2-2s$ which is satisfied,
 with $0\leq s\leq 0.5$. 
 All this can be generalized, that is, if we
have $p$ anyons with 
distinct spins $s$, we obtain
   
\begin{equation}
h_{s}=2-2\sum_{i=1}^{p}s_{i}
\end{equation}
\noindent and
\begin{equation}
{\cal W}_{s}=h_{s}+2\sum_{i=1}^{p}s_{i}-2p,
\end{equation}

\noindent where $p$ is the number of particles, $s_{i}$
is the spin of the
$p_{i}$ particle and $h_{s}$ is the global fractal
properties of the 
set of particles. In any case, ${\cal W}_{s}={\cal W}$,
that is, we have a topological invariant characterizing
 the multiply
connected spaces,  in which each hole represents
 a particle-flux 
of the anyonic model.

Finally, a matter which deserves serious attention 
is the number $h_{s}$,
 that probably contains some information about the 
 interaction between the particles,
  the equivalence classes of the path swept out just by 
  these particles and so on. Further considerations 
  along these lines will be
  published shortly.
 
\acknowledgments
I would like to thank A. E. Gon\c{c}alves for 
useful discussions and 
Steven F. Durrant for reading the manuscript.

\end{document}